\documentstyle[preprint,prd,epsf,aps,12pt]{revtex}
\def\NPB{{\em Nucl. Phys.} {\bf B}}

\def\PRL{\em Phys. Rev. Lett.}

\def\rt{\rightarrow}
\def\ffg{H \rightarrow f \bar{f} \gamma}

\def\be{\begin{eqnarray}}
\def\en{\end{eqnarray}}
\begin{document} 
\thispagestyle{empty}
\preprint{\vbox{
\halign{&##\hfil\cr
        & PUTP-TH-98-01 \cr
        & hep-ph/9801334 \cr
        & January 1998 \cr
        & {} \cr }}}
\title {Radiative Higgs Boson Decays $H\rt f\bar f \gamma$ Beyond the
Standard Model}

\author{Chong-Sheng Li$^{a, b}$, Shou-Hua Zhu$^{a, b}$ and 
Cong-Feng Qiao$^{b}$\\[0.2cm]}

\address{ $^a$ Department of Physics, Peking University, 
Beijing 100871, P.R. China.\\
$^{b}$ CCAST (World Laboratory), P.O. Box 8730, Beijing 100080, P.R.}
\maketitle
\vspace{0.5cm}

\begin{abstract} 

Neutral Higgs boson radiative decays of the form $h_0, H, A\rt
f\bar{f}\gamma$, in the light fermion limit $m_f\rt 0 $,
are calculated in the two Higgs doublet model at one-loop
level. Comparisons with the calculation within the standard model are
given, which indicates that these two models are distinguishable
in the decay mode fermion-antifermion-photon. Our results show that
the concerned process may stand as an implement to identify the
Higgs belongings in case there is a intermediate mass Higgs detected.
\vskip 0.5cm

\noindent
PACS Number(s): 13.85.Qk, 14.70.Hp, 14.80.Bn, 14.80.Cp           

\end{abstract}
\newpage

\section{Introduction}

The discovery of the Higgs boson is one of the most important goals for 
future high energy physics. 
Although the Higgs mass cannot be precisely predicted in the Standard Model  
(SM), it can be constrained and deduced from detecting some processes it  
involves in.  The direct search in the LEP experiments via the  
$e^+ e^- \rt Z^* H$ yields a lower bound of $\sim 77.1$ GeV on the Higgs mass  
\cite{a0}.  This search is being extended at present LEP2 experiments, which  
will probe Higgs masses up to about 95 GeV \cite{a01}. After LEP2 the search  
for the Higgs particles will be continued at the LHC for Higgs masses up to the 
theoretical upper limit. The detection of the Higgs boson at the LHC will 
be divided into two mass region: $M_W \le M_H \le 130$ GeV, the so-called 
intermediate mass range, and $130$ GeV $\le M_H \le 800$ GeV.  
For searching the intermediate mass Higgs boson, the decay $H\rt
\gamma\gamma$ is still one of the important discovery mode \cite{a01d},
although techniques of secondary vertex detection in experiment is  
greatly improved in recent years, which allow the detection of secondary  
vertices from the decay of $b$ quarks in the decay of $H\rt b \bar{b}$ 
at hadron colliders \cite{a1}. For $M_H > 2M_W, 2M_Z$, Higgs decays to
real weak bosons become dominant. 
 
Recent investigation \cite{a2} indicates that the radiative process $\ffg$ 
also has some unique characters and could be used to supplement Higgs 
boson searches for the intermediate mass Higgs boson,  
where $f$ is a light fermion. However, if the Higgs boson should be
detected, to determining whether it is a Higgs boson of the SM or one of
its extensions is also necessary. Many extensions of the SM contain more
than one Higgs doublet. The two Higgs doublet model(2HDM) is one of the
extensions \cite{a3}, which has drawn much attentions these years, because
in the minimal supersymmetric extension of the SM (MSSM) \cite{a3,a4} two
Higgs doublets have to be introduced \cite{a6}. In the 2HDM, there are
three neutral and two charged Higgs bosons, $h_0, H, A, H^{\pm}$ of which
$h_0$ and $H$ are CP-even and $A$ is CP-odd.  
 
In this paper we study the $\ffg$ process in the context of the  
MSSM Higgs sector, where $H$ denotes $h_0, H$, and $A$.  
We present the decay widths versus Higgs mass changing in 
intermediate mass region, and compare with the results of the 
same process in the SM in the case of different parameter choices.  
In the section II, we present expressions for the decay amplitudes.  
In the section III, we give our numerical results and discussions.  
 
\section{FORMALISM OF HIGGS BOSONS DECAY WIDTHS} 
 
The Higgs bosons, $h_0, H, A$ couple to fermions proportionally to their 
masses. Hence, the lowest order diagrams of the processes $h_0, H, A\rt 
f\bar{f} \gamma$ are those of loop diagrams in the $m_f \rt 0$ limit.  
We perform the calculations in the Feynman--'t Hooft gauge 
and generally set $m_f = 0$ except for phase space intergrations.  
 
The relevant Feynman diagrams for those processes are shown in  Fig. 1.  
The amplitudes for $A\rt f\bar{f} \gamma$ can be expressed as  
 
\begin{eqnarray} 
M_{A}=M_{A}^{\gamma}+M_{A}^{Z}, 
\end{eqnarray} 
where 
\begin{eqnarray} 
M_{A}^{\gamma}= 
f_{A}^{\gamma}(k_1\cdot k_3-k_2\cdot k_3) \bar{u}(k_1)~\rlap/\epsilon 
~\gamma_5 v(k_2) 
(k_2-k_1)\cdot\epsilon \bar{u}(k_1)~\rlap/\epsilon~ \gamma_5 v(k_2) 
\end{eqnarray} 
and
\begin{eqnarray} 
M_{A}^{Z}&=&2 f_{A}^{Z}[ 
a_f (k_2\cdot k_3-k_1\cdot k_3) 
\bar{u}\rlap/\epsilon v(k_2) 
- v_f (k_2\cdot k_3-k_1\cdot k_3) 
\bar{u}(k_1)~\rlap/\epsilon~\gamma_5 v(k_2)\nonumber \\ 
&+& a_f
\bar{u}(k_1)\rlap/k_3 \gamma_5 v(k_2)(k_1-k_2)\cdot \epsilon
- v_f
\bar{u}(k_1)\rlap/k_3 \gamma_5 v(k_2)(k_1-k_2)\cdot \epsilon] 
\end{eqnarray} 
with 
\begin{eqnarray} 
f_{A}^{\gamma}&=& {-i e^3 g Q_f  \over 24 (k_1\cdot k_2+m_f^2) m_W \pi^2} 
[\tan\beta m_b^2 C_0(2 k_1\cdot k_2, 0, m_A^2, m_b^2, m_b^2, m_b^2)\nonumber \\ 
&+& 
4 \cot\beta m_t^2 C_0(2 k_1\cdot k_2, 0, m_A^2, m_t^2, m_t^2, m_t^2)],  
\end{eqnarray} 
\begin{eqnarray} 
f_{A}^{Z}&=&{-i e^3 g  \over 
16 \pi^2 \sin\theta_w \cos\theta_w m_W (2 k_1\cdot k_2-m_Z^2+ i \Gamma_Z 
m_Z)} 
\times\nonumber \\ 
&&[(-1/2+ 2/3 \sin^2\theta_w) m_b^2 \tan\beta  
C_0(2 k_1\cdot k_2, 0, m_A^2, m_b^2, m_b^2, m_b^2)\nonumber \\ 
&+& 
2 (-1/2+ 4/3 \sin^2\theta_w) 
\cot\beta m_t^2 C_0(2 k_1\cdot k_2, 0, m_A^2, m_t^2, m_t^2, m_t^2)]. 
\end{eqnarray} 
Here and below, $v_f={I^3_{w,f}-2 \sin^2\theta_w Q_f \over  
2\sin\theta_w \cos\theta_w}$, $a_f= {I^3_{w,f} \over
2\sin\theta_w\cos\theta_w}$,
$k_1, k_2$, and $k_3$ denote momentums of light fermions  
and photon, respectively, $tan\beta=v_2/v_1$, i.e., the ratio of the two   
vacuum expectation values, and $C_0, C_{ij}$ and $D_0, D_{ij}$ are the  
three-point and four-point Feynman integrals \cite{veltman}. 
 
The amplitude for $h_0\rt f\bar{f} \gamma$ is given by 
 
\begin{eqnarray} 
M_{h_0}=M_{h_0}^{tri}+M_{h_0}^{box}, 
\end{eqnarray} 
where 
\begin{eqnarray} 
M_{h_0}^{tri}=M_{h_0}^{tri,\gamma} + M_{h_0}^{tri, Z},~ 
M_{h_0}^{box}=M_{h_0}^{box, W} + M_{h_0}^{box, Z}, 
\end{eqnarray} 
with 
\begin{eqnarray} 
M_{h_0}^{tri, \gamma}&=& M_{h_0}^{tri, \gamma, fermions} + 
M_{h_0}^{tri, \gamma, H^\pm} + 
M_{h_0}^{tri, \gamma, X},\\ 
M_{h_0}^{tri, Z}&=&M_{h_0}^{tri, Z, fermions}+M_{h_0}^{tri, Z, X}+ 
M_{h_0}^{tri, Z, H^\pm},\\ 
M_{h_0}^{box, W}&=&\bar{u}(k_1) 
(\rlap/\epsilon f_1^{box, W} 
+\rlap/k_3 f_2^{box, W} \epsilon\cdot k_1 
+\rlap/k_3 f_3^{box, W} \epsilon\cdot k_2) 
(1-\gamma_5) v(k_2),\\ 
M_{h_0}^{box, Z}&=&\bar{u}(k_1) 
(\rlap/\epsilon f_1^{box, Z} 
+\rlap/k_3 f_2^{box, Z} \epsilon\cdot k_1 
+\rlap/k_3 f_3^{box, Z} \epsilon\cdot k_2) 
[4 a_f^2- {2 I^3_{w,f} Q_f\over \cos^2\theta_w}\nonumber\\ 
 &+&2 (tan\theta_w Q_f)^2 
-4 a_f
(a_f-tan\theta_w Q_f )\gamma_5] v(k_2). 
\end{eqnarray} 
Here, $X$ denotes $W^\pm, G^\pm, \eta^\pm$, and  
\begin{eqnarray} 
M_{h_0}^{tri, \gamma, fermions}&=& 
{e^3 g Q_f \over 24  (k_1\cdot k_2+m_f^2) m_W \pi^2} 
\bar{u}(k_1)[-k_3\cdot (k_1+k_2)\rlap/\epsilon  
+ \rlap/k_3 \epsilon\cdot (k_1+k_2)]v(k_2) 
\nonumber \\ 
&&[(- C_0+4 C_{12})(2 k_1\cdot k_2, 0, m_{h_0}^2, m_b^2, m_b^2, m_b^2) 
\sin\alpha \sec \beta\nonumber\\ 
&-& 4 (- C_0+4 C_{12})(2 k_1\cdot k_2, 0, m_{h_0}^2, m_t^2, m_t^2, m_t^2) 
\cos\alpha \csc\beta], 
\end{eqnarray} 
\begin{eqnarray} 
M_{h_0}^{tri, \gamma, H^\pm}&=&{e^3 g Q_f \over 8 (k_1\cdot k_2+m_f^2) m_W 
\cos\theta_w \pi^2} 
\bar{u}(k_1)[-k_3\cdot (k_1+k_2)\rlap/\epsilon  
+ \rlap/k_3 \epsilon\cdot (k_1+k_2)]v(k_2) 
\nonumber \\ 
&&[-2 \cos\theta_w m_W\sin(\alpha-\beta)+ 
m_Z \cos(2\beta) \sin(\alpha+\beta)] \times\nonumber\\ 
&&C_{12}(2 k_1.k_2, 0, m_{H^\pm}^2, 
m_{H^\pm}^2, m_{H^\pm}^2), 
\end{eqnarray}  
\begin{eqnarray} 
M_{h_0}^{tri, \gamma,X}&=& {-e^3 g Q_f \over 8 (k_1.k_2+m_f^2)\cos\theta_w 
  \pi^2} 
\bar{u}(k_1)[-k_3\cdot (k_1+k_2)\rlap/\epsilon f_{h_0,\gamma}^2 
+ \rlap/k_3 \epsilon\cdot (k_1+k_2)f_{h_0,\gamma}^1]v(k_2) 
\end{eqnarray} 
\begin{eqnarray} 
M_{h_0}^{tri, Z, fermions}&=& 
{-e^3 g \over 96 \cos\theta_w m_W (2 k_1\cdot k_2-m_Z^2+i\Gamma_z m_Z) 
\pi^2 \sin\theta_w} 
\nonumber \\ 
&&\bar{u}(k_1)\{2 v_f[ 
-k_3\cdot (k_1+k_2)\rlap/\epsilon+ \rlap/k_3 \epsilon\cdot (k_1+k_2)] 
\nonumber\\ 
&-&2 a_f
[-k_3\cdot (k_1+k_2)\rlap/\epsilon \gamma_5+ \rlap/k_3 \gamma_5  
\epsilon\cdot (k_1+k_2)]\}  
v(k_2) \nonumber \\ 
&& [\sec\beta \sin\alpha m_b^2 (-3+4 \sin^2\theta_w) 
(C_0-4 C_{12})(2 k_1\cdot k_2, 0, m_{h_0}^2, m_b^2, m_b^2, m_b^2)\nonumber 
\\ 
&-& 
2\csc\beta \cos\alpha m_t^2 (-3+8 \sin^2\theta_w) 
(C_0- 4 C_{12})(2 k_1\cdot k_2, 0, m_{h_0}^2, m_t^2. m_t^2, m_t^2)], 
\end{eqnarray} 
\begin{eqnarray} 
M_{h_0}^{tri, Z, H^\pm}&=& 
{-e^2 g^2 \cos(2 \theta_w)  
\over 8 \cos\theta_w  (2 k_1\cdot k_2-m_Z^2+i\Gamma_z m_Z) \pi^2  
} 
\nonumber \\ 
&&\bar{u}(k_1)\{2 v_f[ 
-k_3\cdot (k_1+k_2)\rlap/\epsilon+ \rlap/k_3 \epsilon\cdot (k_1+k_2)] 
\nonumber\\ 
&-&2 a_f[ -k_3.(k_1+k_2)\rlap/\epsilon \gamma_5+ \rlap/k_3 \gamma_5  
\epsilon\cdot (k_1+k_2)]\}  
v(k_2) \nonumber \\ 
&&[-\sin(\alpha-\beta) m_W +{m_Z \cos(2\beta) \sin(\alpha+\beta) \over 
2 \cos\theta_w}] C_{12}(2 k_1.k_2, 0, m_{H^\pm}^2, 
 m_{H^\pm}^2, m_{H^\pm}^2), 
\end{eqnarray} 
\begin{eqnarray} 
M_{h_0}^{tri, Z, X}&=& 
{-e^2 g^2 \over 32 \cos^2\theta_w  (2 k_1\cdot k_2-m_Z^2+i\Gamma_z m_Z) 
\pi^2  } 
\nonumber \\ 
&&\bar{u}(k_1)\{2 v_f [ 
-k_3\cdot (k_1+k_2) f_{h_0, Z}^2\rlap/\epsilon+ \rlap/k_3 
\epsilon\cdot (k_1+k_2)] 
f_{h_0, Z}^1 \nonumber\\ 
&-& 2 a_f[ 
-k_3\cdot (k_1+k_2)f_{h_0, Z}^2\rlap/\epsilon \gamma_5+ \rlap/k_3 \gamma_5  
\epsilon.(k_1+k_2) f_{h_0, Z}^1]\}
v(k_2). 
\end{eqnarray} 
In above eqs. $f_{h_0,\gamma}^i$, $f_{h_0, Z}^i$,  
$f^{box,W}_i$, and $f^{box,Z}_i$ are form factors, and their explicit  
expressions are given by 
\begin{eqnarray} 
f_{h_0,\gamma}^1&=& -4 \cos\theta_W m_W \sin(\alpha-\beta) C_0
(2 k_1\cdot k_2, 0, m_W^2, m_W^2, m_W^2)+ 
[6 \cos\theta_W m_W \sin(\alpha-\beta) 
\nonumber \\ 
&+& m_Z \cos(2\beta) \sin(\alpha+\beta)] C_{12} 
(2 k_1\cdot k_2, 0, m_W^2, m_W^2, m_W^2), 
\end{eqnarray} 
\begin{eqnarray} 
f_{h_0,\gamma}^2&=& f_{h_0,\gamma}^1 - 
{\cos\theta_w m_W\over 4}[6 \sin(\alpha-\beta)- 
{2 m_{h_0}^2\sin(\alpha-\beta) -m_Z^2\cos(2\beta) \sin(\alpha+\beta) \over 
k_1\cdot k_3+k_2\cdot k_3}]\nonumber \\ 
&&C_0(2 k_1\cdot k_2, 0, m_W^2, m_W^2, m_W^2), 
\end{eqnarray} 
\begin{eqnarray} 
f_{h_0, Z}^1 &=& 
4 \cos^2\theta_w \sin(\alpha-\beta) 
\{(3 \cos\theta_w m_W- m_Z \sin^2\theta_w)C_0
(2 k_1\cdot k_2, 0, m_W^2, m_W^2, m_W^2) 
\nonumber \\ 
&+& 
[- 20  \cos^3\theta_w m_W \sin(\alpha-\beta)+ 
\cos\theta_w\sin^2\theta_w m_W  \sin(\alpha-\beta)+ 
3 \cos^2\theta_w\sin^2\theta_w m_Z\nonumber\\ 
&-&\cos(2\theta_w) m_Z \cos(2\beta) \sin(\alpha+\beta)]C_{12} 
(2 k_1\cdot k_2, 0, m_W^2, m_W^2, m_W^2)\}, 
\end{eqnarray} 
\begin{eqnarray} 
f_{h_0,Z}^2&=& f_{h_0,Z}^1 -{\cos\theta_w m_W\over 4} 
C_0(2 k_1\cdot k_2, 0, m_W^2, m_W^2, m_W^2) 
\{6 \sin(\alpha-\beta)\cos^2\theta_w \nonumber\\ 
&-& 
{[(2\cos^2\theta_w +1)m_{h_0}^2-3\cos^2\theta_w m_Z^2]\sin(\alpha-\beta) 
-m_Z^2\sin^2\theta_w \cos(2\beta) \sin(\alpha+\beta) \over 
k_1\cdot k_3+k_2\cdot k_3}\}, 
\end{eqnarray} 
\begin{eqnarray} 
f_1^{box, W}&=& 
{e g^3 m_W \sin(\alpha-\beta)\over 128 \pi^2} 
\{3 C_0(0, 2 k_2\cdot k_3, m_{h_0}^2, m_W^2,0,m_W^2)\nonumber\\
&+& 
(3 m_W^2-8  k_2\cdot k_3) D_0 (0, 0, m_{h_0}^2,0, 2 k_1\cdot k_2, 2
k_2\cdot k_3, m_W^2,0, m_W^2,m_W^2)
\nonumber\\ 
&-& 8 k_2\cdot k_3 D_1 (0, 0, m_{h_0}^2,0, 2 k_1\cdot k_2, 2 k_2\cdot k_3,
m_W^2,0, m_W^2,m_W^2)\nonumber\\
&+&
2 (3 k_1\cdot k_2+4 k_1\cdot k_3-4 k_2\cdot k_3)D_2 
(0, 0, m_{h_0}^2,0, 2 k_1\cdot k_2, 2 k_2\cdot k_3, m_W^2,0, 
m_W^2,m_W^2)
\nonumber \\ 
&-&6 k_2\cdot k_3 D_3 
(0, 0, m_{h_0}^2,0, 2 k_1\cdot k_2, 2 k_2\cdot k_3, m_W^2,0, 
m_W^2,m_W^2)\nonumber\\
&+&
8 D_{00}(0, 0, m_{h_0}^2,0, 2 k_1\cdot k_2, 2 k_2\cdot k_3, m_W^2,0, 
m_W^2,m_W^2)
\}\nonumber\\ 
&+& (k_1\cdot k_3 \leftrightarrow k_2\cdot k_3), 
\end{eqnarray} 
\begin{eqnarray} 
f_2^{box, W}&=& 
-{e g^3 m_W \sin(\alpha-\beta) \over 16 \pi^2} 
(D_1+D_{23})(0, 0, m_{h_0}^2,0, 2 k_1\cdot k_2, 2 k_2\cdot k_3, m_W^2,0, 
m_W^2,m_W^2)
\nonumber \\ 
&+&(k_1\cdot k_3 \leftrightarrow k_2\cdot k_3), 
\end{eqnarray} 
\begin{eqnarray} 
f_3^{box, W}&=& 
{e g^3 m_W \sin(\alpha-\beta) \over 16 \pi^2} 
(D_0+D_2 +D_2- D_{13}-D_{23})(0, 0, m_{h_0}^2,0, 2 k_1\cdot k_2, 2
k_2\cdot k_3, m_W^2,0,m_W^2,m_W^2)\nonumber \\ 
&+&(k_1\cdot k_3 \leftrightarrow k_2\cdot k_3), 
\end{eqnarray} 
\begin{eqnarray} 
f_1^{box, Z}&=& 
{e^3 g m_Z \sin(\alpha-\beta) Q_f \over 16 \pi^2 \cos\theta_w} 
[-C_0(0,k_2\cdot k_3, m_{h_0}^2,m_Z^2,m_f^2,m_Z^2) \nonumber \\  
&+&2( k_2\cdot k_3 D_1  
+D_{00})(0, 0, m_{h_0}^2,0, 2 k_1\cdot k_2, 2 k_2\cdot k_3, m_f^2,m_Z^2, 
m_Z^2,m_f^2)], 
\end{eqnarray} 
\begin{eqnarray} 
f_2^{box, Z}&=& 
{e^3 g m_Z \sin(\alpha-\beta) Q_f\over 8 \pi^2 \cos\theta_w}(D_{22} 
+D_{23})(0, 0, m_{h_0}^2,0, 2 k_1\cdot k_2, 2 k_2\cdot k_3, m_f^2,m_Z^2, 
m_Z^2,m_f^2), 
\end{eqnarray} 
\begin{eqnarray} 
f_3^{box, Z}&=& 
-{e^3 g m_Z \sin(\alpha-\beta) Q_f \over 8 \pi^2 \cos\theta_w} 
( D_1+D_{12}+D_{13})(0, 0, m_{h_0}^2,0, 2 k_1\cdot k_2, 2 k_2\cdot k_3,  
m_f^2,m_Z^2). 
\end{eqnarray} 
The amplitude of the process $H \rightarrow f\bar{f}\gamma$ can be simply 
obtained by substituting $\alpha \rightarrow {3 \pi \over 2} +\alpha $ and  
$m_{h_0} \rightarrow m_{H}$ in the amplitude of $h_0 \rightarrow f\bar{f} 
\gamma$.  

For the simplicity of calculating the amplitude squares, we can  
parameterize the amplitudes of the process $(h_0, H, A)  
\rightarrow f \bar{f}\gamma$ in a general form 
\begin{eqnarray} 
M=\bar{u}(k_1)(g_1 \rlap/\epsilon +g_2 \rlap/\epsilon\gamma_5 
+g_3 \rlap/k_3 \epsilon.k_1+ 
g_4 \rlap/k_3 \gamma_5 \epsilon.k_1+ 
g_5 \rlap/k_3 \epsilon.k_2+ 
g_6 \rlap/k_3 \gamma_5 \epsilon.k_2) 
v(k_2). 
\end{eqnarray} 
Therefore, the amplitude square is given by  
\begin{eqnarray} 
\sum_{spins}|M|^2=8 [(g_1^2+g_2^2) (k_1~\cdot k_2)+ 
 2 Re(g_3 g_5^\dagger+g_4 g_6^\dagger) (k_1\cdot k_2~k_2\cdot k_3
~k_1\cdot k_3)]. 
\end{eqnarray} 
Here $g_i$ are form factors, which can be expressed as the combinations of 
the form factors given above. Their tedious expressions are not shown here. 
 
The differential decay widths can be written as 
\begin{eqnarray} 
{d\Gamma(h_0, H , A \rightarrow f \bar{f} \gamma)\over d(k_1\cdot k_2)}= 
{1\over 256 \pi^3} {1\over m_{h_0,H,A}^3}  
\int_{(k_2\cdot k_3)_{min}}^{(k_2\cdot k_3)_{max}} d(k_2\cdot k_3) 
\sum_{spin} |M|^2 
\nonumber 
\end{eqnarray} 
with 
\begin{eqnarray} 
(k_2 \cdot k_3)_{min}&=& 
{1\over 4}[m_{h_0,H,A}^2-2 (m_f^2+k_1\cdot k_2)](1-\sqrt{1-{2 m_f^2 \over 
 m_f^2+k_1\cdot k_2}}),\nonumber \\ 
(k_2\cdot k_3)_{max}&=& 
{1\over 4}[m_{h_0,H,A}^2-2 (m_f^2+k_1\cdot k_2)](1+\sqrt{1-{2 m_f^2 \over 
m_f^2+k_1\cdot k_2}})\nonumber. 
\end{eqnarray} 
 
\section{NUMERICAL RESULTS AND DISCUSSIONS}  
In our numerical calculation the relevant parameters are chosen as 
\be 
m_t=176~GeV,~ m_b=4.5~GeV, \alpha(M_z)=1/128,\nonumber\\  
M_z=91.2~GeV,~ M_w=80.3~GeV,~\Gamma_z=2.5~GeV, 
\en   
The Higgs boson masses $m_{h_0}$, $m_H$, and $m_{H^\pm}$ are determined
by  $m_A$ and tan$\beta$ as follows \cite{a7} 
\begin{eqnarray} 
m_{h_0}^2&=&{1\over 2}\left[m_A^2 + M_z^2 + \epsilon -
\sqrt{(m_A^2 + M_z^2 + \epsilon)^2- 
4 m_A^2 M_z^2 cos^22\beta - 4\epsilon 
(m_A^2sin^2\beta + M_z^2cos^2\beta)}\right], 
\end{eqnarray}  
\begin{eqnarray} 
m_H^2 = m_A^2 + M_z^2-m_{h_0}^2 + \epsilon,
\end{eqnarray} 
and
\begin{eqnarray}
m_{H^\pm}^2=m_A^2 + m_W^2
\end{eqnarray}
with 
\begin{eqnarray} 
\epsilon =\frac{3 G_F}{\sqrt{2}\pi^2}\frac{m_t^4}{\sin^2\beta} \log (1+ 
\frac{m_S^2}{m_t^2} ). 
\end{eqnarray} 
Here the $m_S$ is a common squark mass which is equal to $1 TeV$ in our
numerical calculations. The mixing angle $\alpha$ is fixed by $\tan\beta$
and the Higgs boson mass $m_A$. 
\begin{eqnarray} 
\tan 2\alpha=\tan 2\beta{m_A^2 + M_z^2 \over m_A^2 - M_z^2 +\epsilon/cos 
2\beta}, 
\end{eqnarray} 
where 
$-{\pi \over 2}<\alpha <0$. 
 
Fig. 2 and Fig.3 show the total decay widths of the processes 
$h_0, H, A\rt f\bar{f} \gamma$ versus the Higgs mass  
varying in the intermediate range for two values $\tan\beta=1.5$  
and $\tan\beta=30$. As can be seen from these figures, the total decay
widths for  $h_0, H, A \rt f\bar{f} \gamma$, where the neutrino,
electron, muon, and light quarks contributions are  
included, have some obvious characters comparing with the two-photon
widths of the MSSM Higgs bosons decays \cite{a6p,a8}.  
First, in the case of tan$\beta=1.5$  the width $\Gamma_{H\rt f\bar{f}
\gamma}$ can exceed the width $\Gamma_{H\rt \gamma\gamma}$ 
for $140 \rm Gev\le m_H\le 200 \rm Gev$. However, in the same Higgs-mass
range $\Gamma_{h_0 \rightarrow f\bar f\gamma}$ and $\Gamma_{A \rightarrow
f\bar f\gamma}$ are smaller than $\Gamma_{h_0 \rightarrow
\gamma \gamma}$ and $\Gamma_{A \rightarrow \gamma \gamma}$, respectively.
And, the width $\Gamma_{H\rt f\bar{f}\gamma}$ less than that of the the
width $\Gamma_{H\rt\gamma\gamma}$ for $m_H< 140 \rm Gev$.

Second, in the case of $\tan\beta=30$ the width $\Gamma_{H \rightarrow
f\bar f\gamma}$ is still larger than $\Gamma_{H \rightarrow \gamma
\gamma}$ for 140 GeV $~< M_H <$ 200 GeV and
$\Gamma_{h_0 \rightarrow f\bar f\gamma}$
is smaller than $\Gamma_{h_0 \rightarrow \gamma \gamma}$.
Only in the vicinity of $M_{h_0,H} \approx 130$ GeV
the decay widths $\Gamma_{h_0 \rightarrow f\bar f\gamma}$ and 
$\Gamma_{H \rightarrow f\bar f\gamma}$
are about the same as $\Gamma_{h_0 \rightarrow \gamma \gamma}$
and $\Gamma_{H \rightarrow \gamma \gamma}$, respectively.
Again, the width for the radiative decay of the pseudoscalar
$\Gamma_{A \rightarrow f \bar f \gamma}$
is smaller than $\Gamma_{A \rightarrow \gamma \gamma}$
for $M_A<200$ GeV.
 
Comparing with the same process of the SM \cite{a2},  
we find that in general the widths in the MSSM case are less than that in
the SM case, except Higgs mass  is around $130$ Gev, where the predictions
of the SM and MSSM on the $f\bar{f}\gamma$ widths are almost the same
and indistinguishable.  
 
In conclusion, in addition to be a supplement in searching the Higgs
boson through the Higgs decay to two photons, the radiative decay 
of Higgs boson $H\rt f\bar{f}\gamma$ would be an more observable channel 
in searching the Higgs boson in the future experiment since the radiative 
decay widths can be larger than the two-photon decay mode for some
favorable parameter space. Besides, our calculation also shows that
the process $h_0, H, A\rt f\bar{f}\gamma$  may play an important role in
identifying the Higgs boson being of the SM or the MSSM on the basis of
the size of decay widths if the Higgs boson is discoved via this process.  
 
\vskip 1cm 
\begin{center} 
\bf\large\bf{Acknowledgements} 
\end{center} 
This work was supported in part by the National Natural Science Foundation 
of China, the State Commission of Science and Technology of China, and the
Hua Run Postdoctoral Science Foundation of China.

\newpage 
\centerline{\bf \large Figure Captions} 
\vskip 2cm 
\noindent 
Fig.1. The generic Feynman diagrams of $h_0, H, A \rt f\bar{f} \gamma$ 
processes. 
 
\noindent 
Fig.2. The neutral Higgs decay widths versus Higgs masses of $h_0, H, A 
\rt f\bar{f} \gamma$ processes in MSSM with tan$\beta=1.5$.

\noindent 
Fig.3. The neutral Higgs decay widths versus Higgs masses of $h_0, H, A 
\rt f\bar{f} \gamma$ processes in MSSM with tan$\beta=30$.

\newpage
\begin{figure}
\epsfxsize=15 cm
\centerline{\epsffile{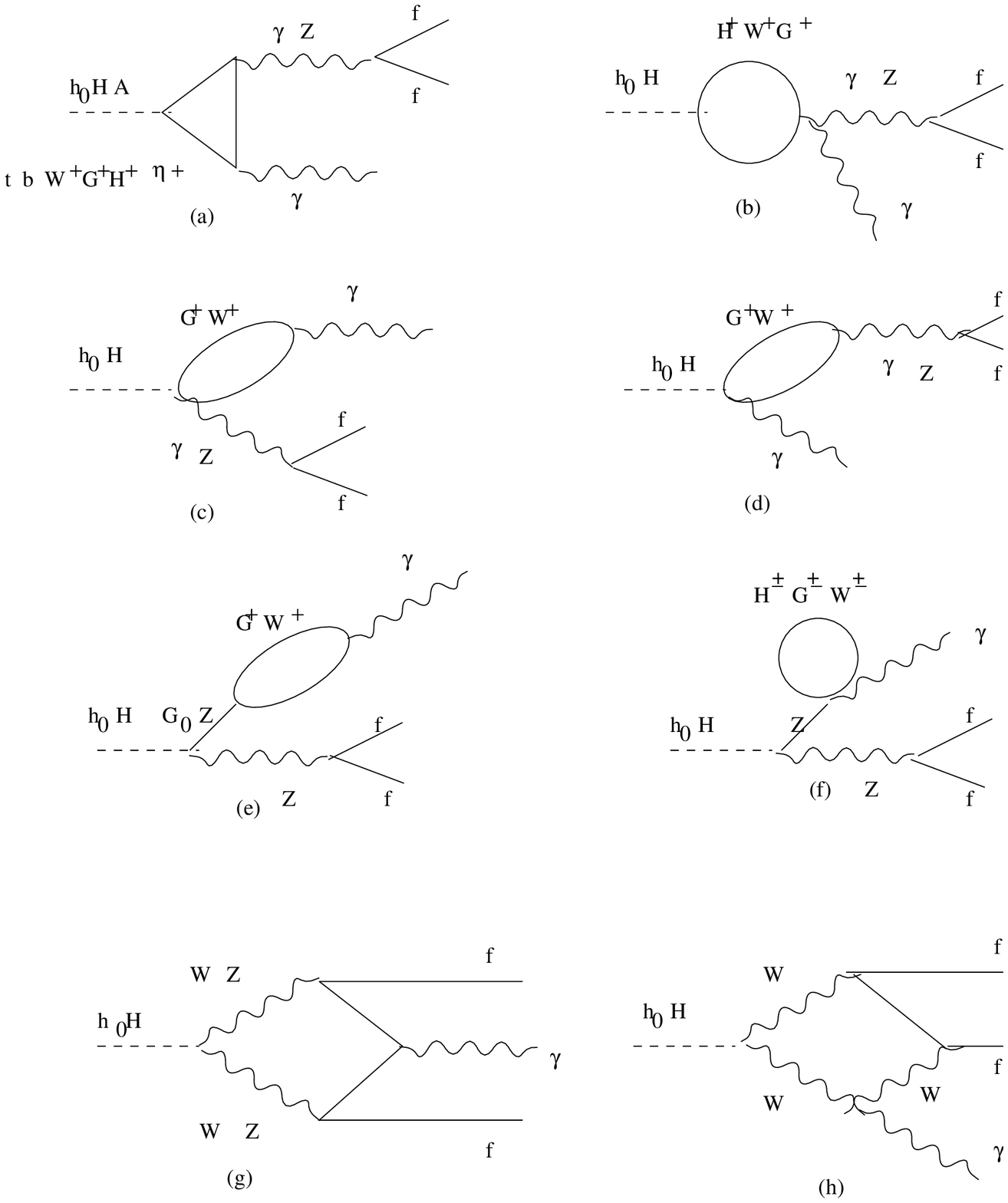}}
\caption[]{}
\end{figure}

\begin{figure}
\epsfxsize=15 cm
\centerline{\epsffile{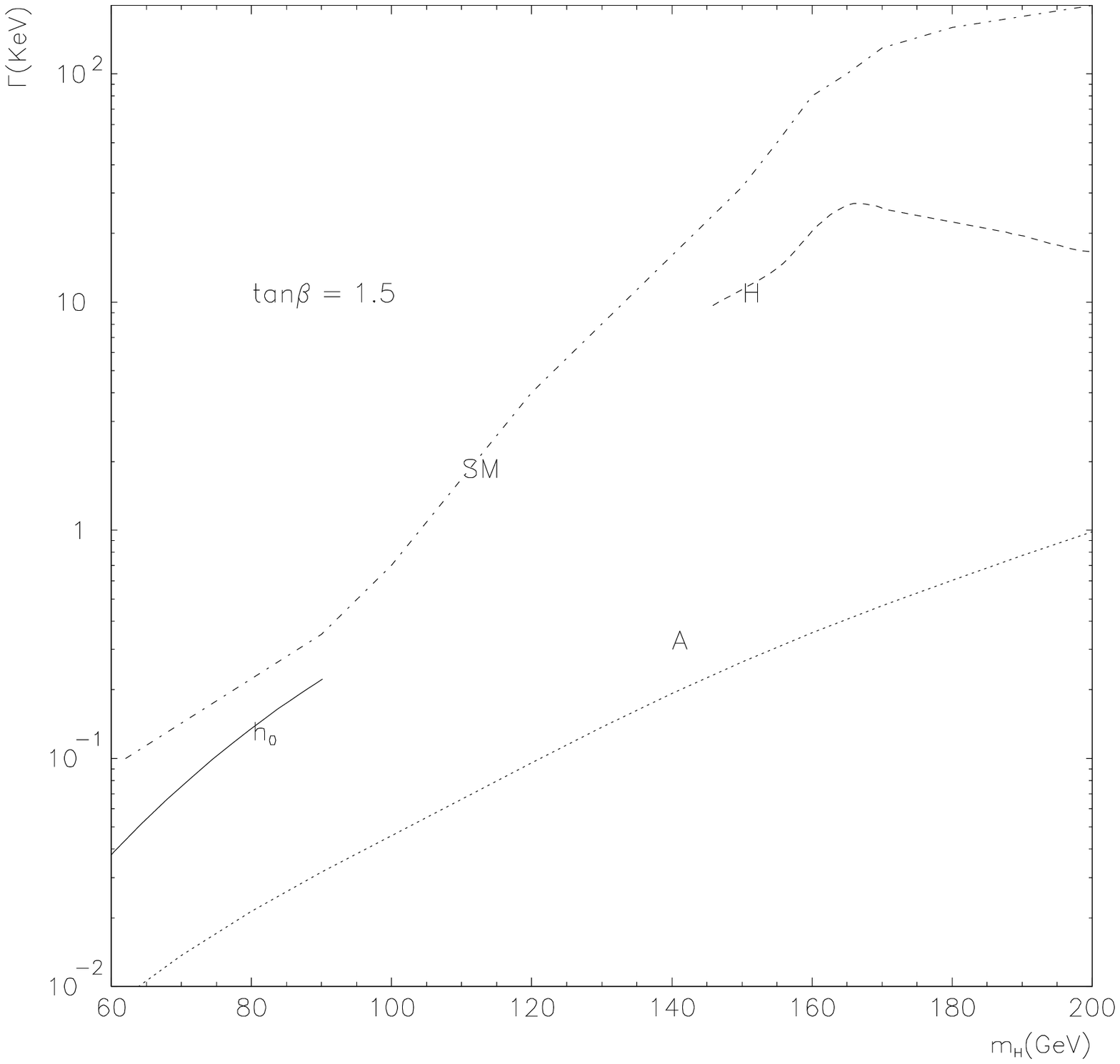}}
\caption[]{}
\end{figure}

\begin{figure}
\epsfxsize=15 cm
\centerline{\epsffile{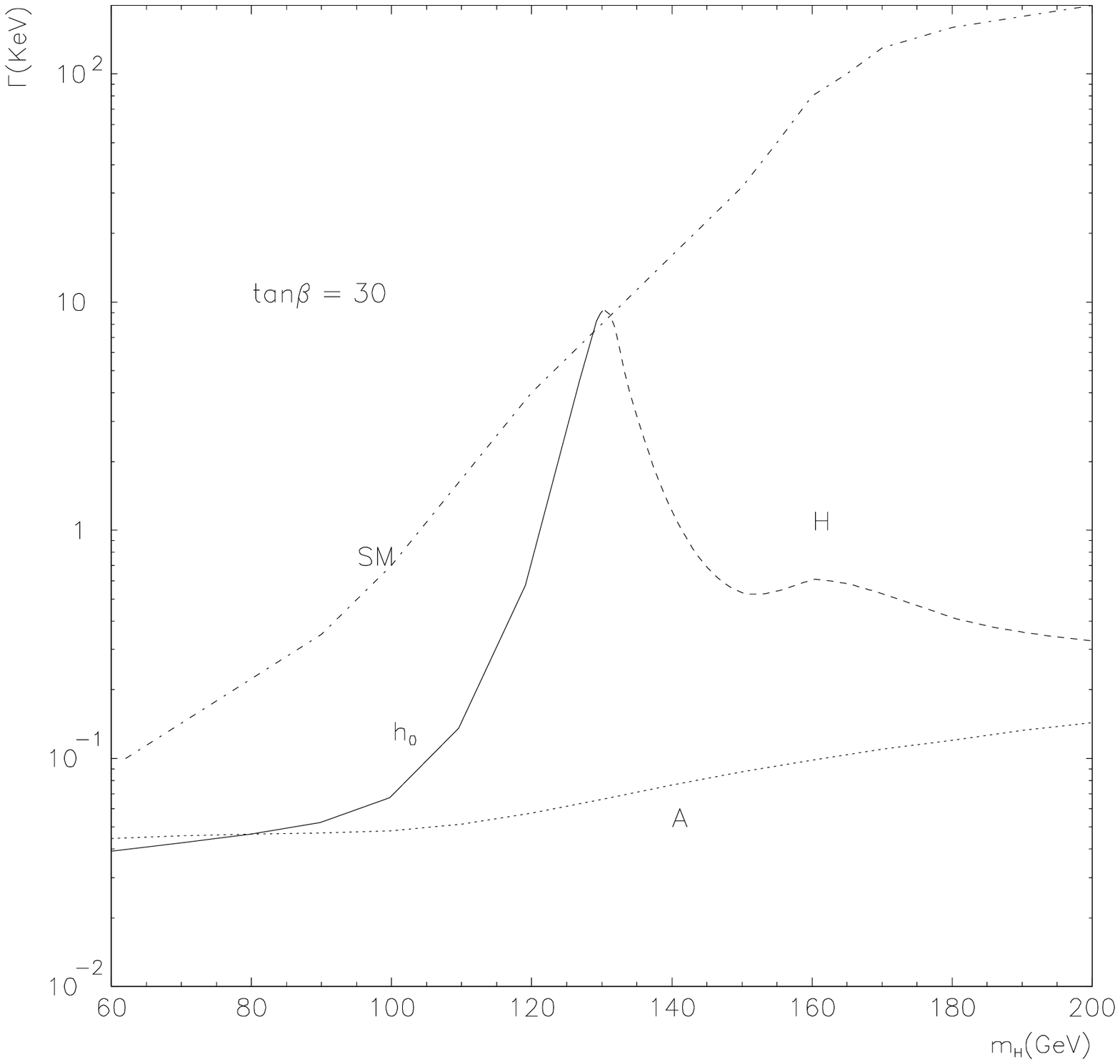}}
\caption[]{}
\end{figure}

\end{document}